\documentclass[acmtog]{acmart}
\AtBeginDocument{%
  }

\copyrightyear{2024}
\acmYear{2024}
\setcopyright{acmlicensed}\acmConference[SA Conference Papers '24]{SIGGRAPH Asia 2024 Conference Papers}{December 3--6, 2024}{Tokyo, Japan}
\acmBooktitle{SIGGRAPH Asia 2024 Conference Papers (SA Conference Papers '24), December 3--6, 2024, Tokyo, Japan}
\acmDOI{10.1145/3680528.3687677}
\acmISBN{979-8-4007-1131-2/24/12}

\begin{document}

\title{SIGGesture: Generalized Co-Speech Gesture Synthesis via Semantic Injection with Large-Scale Pre-Training Diffusion Models}

\author{Qingrong Cheng}
\email{kirolcheng@tencent.com}
\orcid{0000-0001-6631-1504}
\affiliation{%
  \institution{Tencent AI Lab, Tencent TiMi L1 Studio }
  \city{ShenZhen}
  \country{China}
}
\author{Xu Li}
\email{axuli@tencent.com}
\orcid{0009-0002-1365-6546}
\affiliation{%
  \institution{Tencent AI Lab}
  \city{ShenZhen}
  \country{China}
}
\author{Xinghui Fu}
\email{xinghuifu@tencent.com}
\orcid{0000-0002-7143-0185}
\affiliation{%
  \institution{Tencent AI Lab}
  \city{ShenZhen}
  \country{China}
}

\author{Fei Xia}
\email{wallacexia@tencent.com}
\orcid{0009-0008-5414-7717}
\affiliation{%
  \institution{Tencent TiMi L1 Studio}
  \city{ShenZhen}
  \country{China}
}
\author{Zhongqian Sun}
\email{sallensun@tencent.com}
\orcid{0000-0003-1812-6085}
\affiliation{%
  \institution{Tencent AI Lab}
  \city{ShenZhen}
  \country{China}
}

\renewcommand{\shortauthors}{Qingrong Cheng et al.}

% \begin{abstract}
% The automated synthesis of high-quality 3D gestures from speech is of significant value in virtual humans and gaming. 
% Previous methods focus on synthesizing gestures that are synchronized with speech rhythm, yet they frequently overlook the inclusion of semantic gestures. 
% These are sparse and follow a long-tailed distribution across the gesture sequence, making them difficult to learn in an end-to-end manner.
% Moreover, generating gestures, rhythmically aligned with speech, faces a significant issue that cannot be generalized to in-the-wild speeches. 
% To address these issues, we introduce SIGGesture, a novel diffusion-based approach for synthesizing realistic gestures that are of both high quality and semantically pertinent. 
% Specifically, we firstly build a strong diffusion-based foundation model for rhythmical gesture synthesis by pre-training it on a collected large-scale dataset with pseudo labels.
% Secondly,  we leverage the powerful generalization capabilities of Large Language Models (LLMs) to generate proper semantic gestures for the various speech content. 
% Finally, we propose a semantic injection module to infuse semantic information into the synthesized results during diffusion reverse process. 
% Extensive experiments demonstrate that the proposed SIGGesture significantly outperforms existing baselines and shows excellent generalization and controllability.
% \end{abstract}

\begin{abstract} %--from axuli
The automated synthesis of high-quality 3D gestures from speech holds significant value for virtual humans and gaming.
Previous methods primarily focus on synchronizing gestures with speech rhythm, often neglecting semantic gestures.
These semantic gestures are sparse and follow a long-tailed distribution across the gesture sequence, making them challenging to learn in an end-to-end manner.
Additionally, generating rhythmically aligned gestures that generalize well to in-the-wild speech remains a significant challenge.
To address these issues, we introduce SIGGesture, a novel diffusion-based approach for synthesizing realistic gestures that are both high-quality and semantically pertinent.
Specifically, we firstly build a robust diffusion-based foundation model for rhythmical gesture synthesis by pre-training it on a collected large-scale dataset with pseudo labels.
Secondly,  we leverage the powerful generalization capabilities of Large Language Models (LLMs) to generate appropriate semantic gestures for various speech transcripts. 
Finally, we propose a semantic injection module to infuse semantic information into the synthesized results during the diffusion reverse process. 
Extensive experiments demonstrate that SIGGesture significantly outperforms existing baselines, exhibiting excellent generalization and controllability.
\end{abstract}



%\ccsdesc[500]{Computing methodologies~Animation}
%\ccsdesc[300]{Neural networks}
%\ccsdesc{Do Not Use This Code~Generate the Correct Terms for Your Paper}
%\ccsdesc{Neural Netwroks}
%\ccsdesc[100]{Neural Networks}



\maketitle

\section{Introduction}
\label{sec:intro}

Body gesture can significantly enhance people's expressiveness in communication by emphasizing specific intentions or conveying specific semantics \cite{cassell1999speech,goldin2013gesture,yoon2020speech,nyatsanga2023comprehensive}. Additionally, high-quality 3D gesture animation can present vivid virtual characters that help listeners to concentrate and improve the intimacy between humans and virtual characters. 
Consequently, there is a growing demand for automatic high-quality co-speech gesture synthesis in various applications, including virtual humans, non-player gaming characters, and robot assistants. Many researchers have proposed lots of approaches including rule-based methods \cite{huang2012robot,marsella2013virtual} and data-driven methods \cite{zhi2023livelyspeaker, zhu2023taming, ao2023gesturediffuclip}. 

\begin{figure*}[htb]
  \centering
  \includegraphics[width=0.90\textwidth]{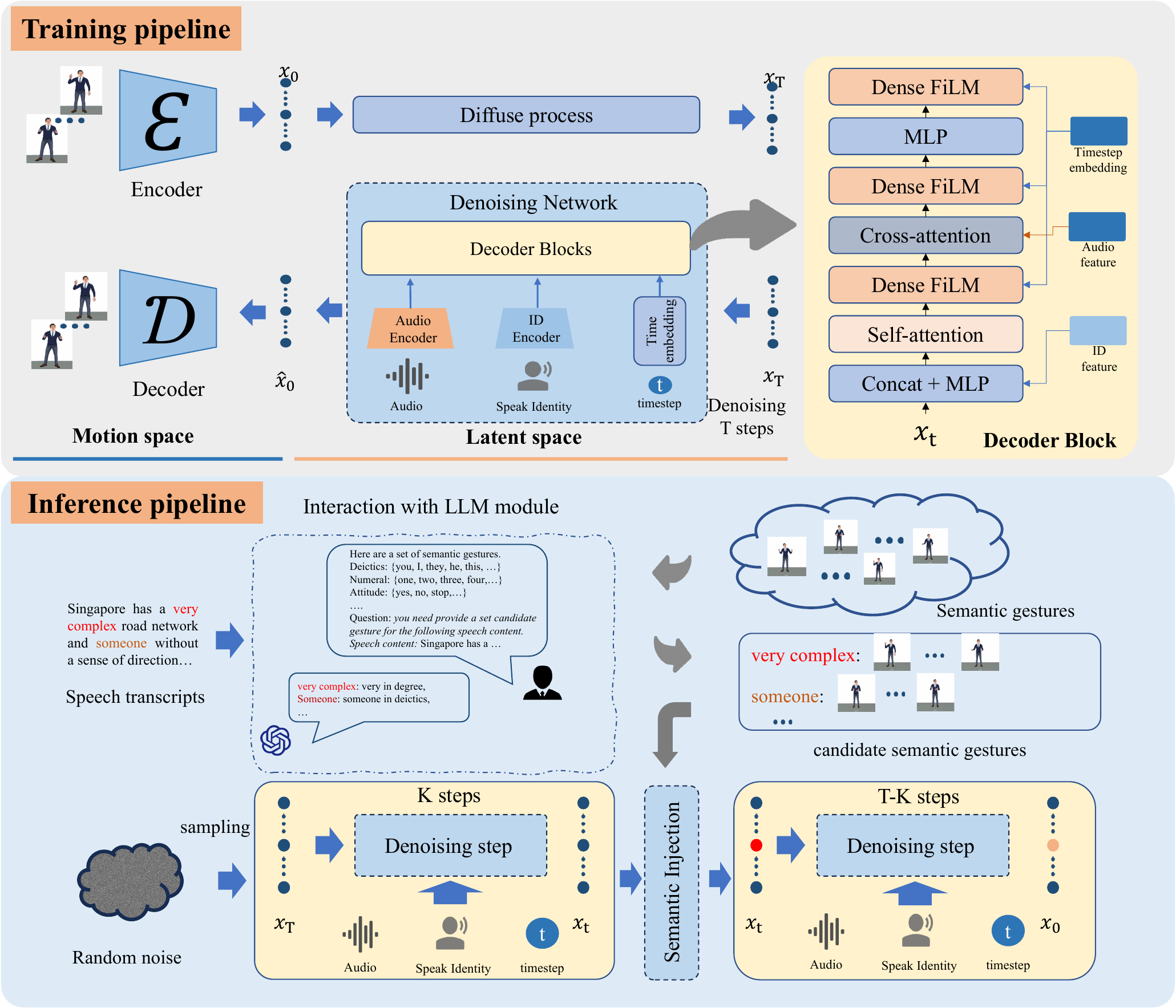}
  \caption{The framework of the proposed method is illustrated in two parts. The upper section details the training processes of diffusion and denoising. The lower part demonstrates the inference process for synthesizing gestures based on the given conditions. Specifically, the audio features and speaker identity are directly fed into the denoising network. Meanwhile, the speech textual content are used as input in the interaction with the LLM, which generates a set of candidate semantic gestures for the subsequent semantic injection process. 
  }
  \label{pipeline}
\end{figure*}

3D co-speech gesture generation aims to produce gestures with natural movements, accurate rhythm, and clear semantics based on the given speech.
Rule-based methods \cite{huang2012robot,marsella2013virtual} achieve this by using predefined rules, such as associating the keyword "good" with a simple "thumbs up" gesture. However, these approaches, lacking naturalness, smoothness, and diversity, can only handle simple semantic gestures.
The main reason is that generating co-speech gestures is challenging due to the inherently complex many-to-many mapping.
Recently, deep learning with large-scale datasets has achieved significant breakthroughs, especially in content synthesis. With the aid of deep learning, co-speech gesture synthesis has seen the emergence of many remarkable methods. These methods attempt to solve this problem by training a conditional model with multi-modal inputs (audio, text, speaker identities, etc.).  

Co-speech gestures include rhythmic gestures and semantic gestures. Some works attempt to generate high-quality co-speech gesture that contains accurate semantic gestures , such as GestureDiffuCLIP \cite{ao2023gesturediffuclip}, LivelySpeaker \cite{zhi2023livelyspeaker}, and Bodyformer \cite{pang2023bodyformer}. Both GestureDiffuCLIP \cite{ao2023gesturediffuclip} and LivelySpeaker \cite{zhi2023livelyspeaker} use CLIP \cite{radford2021learning} to learn the relationship between semantic transcripts and motion. 
However, semantic gestures constitute only a minor portion of the entire sequence and follow a long-tailed distribution, complicating the direct learning of their multi-modal relationships.
Moreover, synthesizing gestures rhythmically synchronized with speech faces the challenge of generalizing to in-the-wild scenarios.

\begin{table*}[tb]
\caption{Comparison of co-speech gesture datasets. We ignore some datasets with 2D and 3D key-points label. "3D Rot" indicates 3D skeleton with bone rotation. "PGT" means Pseudo Ground Truth. }
\label{datasets}
\centering
\resizebox{1.0\linewidth}{!}{
\begin{tabular}{lcccccccc}
\toprule
& \begin{tabular}[c]{@{}c@{}}Trinity\\Mocap 2017\\ \cite{ferstl2018investigating}\end{tabular} & %\begin{tabular}[c]{@{}c@{}}S2G\\S2G-2D\\2019 \cite{ginosar2019learning}\end{tabular} &
%\begin{tabular}[c]{@{}c@{}}Seq2Seq\\TED-2D\\2019 \cite{yoon2019robots}\end{tabular} & 
\begin{tabular}[c]{@{}c@{}}TWH\\Mocap 019\\ \cite{lee2019talking} \end{tabular} & %\begin{tabular}[c]{@{}c@{}}Trimodal \\TED-3D\\2020 \cite{yoon2020speech}\end{tabular}  &
\begin{tabular}[c]{@{}c@{}}Hahibie \textit{et al.} \\S2G-3D 2021\\ \cite{habibie2021learning}\end{tabular} &
%& \begin{tabular}[c]{@{}c@{}}HA2G\\TED-3D+\\2022 \cite{ha2g:liu2022learning}\end{tabular} &
\begin{tabular}[c]{@{}c@{}}BEAT(X)\\Mocap 2022\\ \cite{liu2022beat} \end{tabular} & \begin{tabular}[c]{@{}c@{}}ZEEG\\Mocap 2022\\ \cite{ghorbani2023zeroeggs}\end{tabular} & \begin{tabular}[c]{@{}c@{}}Yoon. \textit{et al.}\\TED-SMPL 2023\\ \cite{lu2023co}\end{tabular} & \begin{tabular}[c]{@{}c@{}}Talkshow\\S2G-SMPL 2023\\ \cite{yi2023talkshow}\end{tabular} & \begin{tabular}[c]{@{}c@{}}\textbf{Gesture400}\\\textbf{Ours}\\\textbf{2024}\end{tabular}  \\ 
\hline
Data type& 3D Rot  & 3D Rot& 3D Rot &3D & 3D Rot& PGT 3D Rot& PGT 3D Rot& PGT 3D Rot \\
Duration (hours)& 4& 20& 38  & 76& 4& 30& 27  & 393\\
\bottomrule
\end{tabular}
}
\end{table*} 

To address the aforementioned issues, we propose a diffusion-based semantic injection method with large language model for semantic gesture synthesis, which offers excellent controllability, interpretability, and generalization. It begins with the insight that real-world human conversation contains relatively limited semantic gestures. For example, for specific semantic words like "wonderful," different people usually present similar gestures, such as "opening both arms." 
During the diffusion reverse process, we can utilize semantically matched gestures as prompts to guide the generative model in synthesizing appropriate semantic gestures. Therefore, we collect a large-scale semantic motion database that includes most semantic motions in communication.
LLMs, such as GPT3 \cite{brown2020language} and GPT-4 \cite{gpt4}, have demonstrated their ability not only as proficient models for natural language processing (NLP) tasks but also as potent instruments for tackling complex problems.  
Therefore, we introduce large-language models to assign contextually appropriate gestures to the corresponding text. 
Leveraging the strong semantic analysis capabilities of LLMs, the proposed method is adept at handling various languages. 

For a specific speech, different people will present notably different rhythmic gestures. For such many-to-many issue, the best solution is to use deep learning with a large-scale dataset to learn their complex relationships. Therefore, it is necessary to build a robust foundational model that can synthesize natural and smooth rhythmic gestures. Then, the semantic gestures can be naturally injected into the rhythmic gestures. 
However, the co-speech gesture generation research community lacks 3D skeletal data due to the expensive motion capture system, which is different from other synthesis tasks, such as text-to-image, text-to-speech. Therefore, for improving the basic quality of rhythmic gestures, we collected a large-scale gesture dataset named Gesture400 for pre-training, which contains approximately 400 hours of data. Table \ref{datasets} shows the statistical data of these co-speech gesture datasets. Some datasets with
2D and 3D key-point labels are ignored because they are difficult to use in real applications. To the best of our knowledge, the proposed dataset is the largest dataset in the field of co-speech gesture synthesis research. By the optimization of pre-training and fine-tuning, the synthesized gesture motion exhibits excellent naturalness and generalization.

The main contributions of our work contain the following aspects:
\begin{enumerate}
	\item We propose an LLMs-based semantic augmentation method for co-speech gesture synthesis, which can greatly improve the generalization ability in semantic gesture synthesis. For controllable semantic gesture synthesis, we build a set of semantic gesture clips and use LLMs to generate appropriate candidate gestures for the specific speech transcripts.  
	
	\item We build a robust diffusion-based foundational model for rhythmical gesture synthesis by pre-training it on a collected large-scale dataset with pseudo labels. This dataset is the largest dataset for co-speech gesture synthesis, containing approximately 400 hours of motion sequences. 
	
	\item Extensive experiments show that the proposed method outperforms state-of-the-art methods by a large margin. In particular, the visualization comparisons indicate that our method produces more stable, expressive, and robust results than other approaches. 
 
\end{enumerate}

\section{Related Work}
\subsection{Co-Speech Gesture Generation}
%traditional
%rencent (leaning based, diffusion)
The research of co-speech gesture generation can be divided into two branches, rule-based methods \cite{cassell1994animated, marsella2013virtual,cassell2001beat} and learning-based methods \cite{ao2022rhythmic,ao2023gesturediffuclip,liu2022learning,ginosar2019learning}. 
Rule-based works \cite{cassell1994animated, marsella2013virtual,cassell2001beat} for co-speech gesture synthesis are mainly generated from a pre-defined motion database by keyword matching or other specific rules. These rule-based methods require lots of human effort in defining gesture units and complex mapping rules, which are costly and inefficient. Besides, the results of rule-based methods are usually lack of smoothness and naturalness. 

Co-speech gesture synthesis is a complex problem that requires consideration of the audio, the gesture, and their many-to-many relationships. Benefit from recent advanced developments of deep learning, all of recent methods \cite{ao2022rhythmic,ao2023gesturediffuclip,liu2022learning} adopt deep neural networks as a strong tool to learn the complex relationships from audio to gesture in an end-to-end manner. Earlier  works \cite{ginosar2019learning,qian2021speech} explore various network architecture to regress 2D keypoints of the human body movements because of lacking high-quality 3D gesture datasets. These 2D datasets are constructed by using off-the-shelf pose estimator \cite{cao2017realtime} to obtain pseudo 2D gesture annotations from the online videos. However, The 2D results face significant challenges in practical applications.. 
Recently, some high-quality speech-gesture corpus such as BEAT \cite{liu2022beat} and ZeroEGGS \cite{ghorbani2023zeroeggs} are released, which contribute to the development of co-speech gesture synthesis. Some of existing works focus on exploring the effectiveness of different network architectures, including the Multi-Layer Perceptron (MLP) \cite{kucherenko2020gesticulator}, Convolutional Neural Networks (CNN) \cite{habibie2021learning,yi2023talkshow}, Recurrent Neural Networks (RNNs) \cite{yoon2020speech,hasegawa2018evaluation,liu2022learning,bhattacharya2021speech2affectivegestures} and Transformers \cite{bhattacharya2021text2gestures,pang2023bodyformer}. Besides the exploration of model architecture, some approaches \cite{ginosar2019learning,ijcai2023DiffuseStyleGesture} dive into researching the connections between co-speech gesture and speech audio, text transcript, speaking style, and speaker identity. 
Furthermore, some approaches involve the adversarial training \cite{ginosar2019learning,liu2022learning}, phase-guided motion matching\cite{yang2023qpgesture}, and reinforcement learning \cite{sun2023co} to guarantee realistic results. 
In addition to gestures, some methods \cite{yi2023talkshow, chen2024diffsheg,liu2023emage} begin to explore the generation of full-body movements, including body gestures and facial expressions.

The rapid development of diffusion models \cite{ho2020denoising,song2020denoising} has surpassed the traditional synthesis paradigm such as GAN \cite{goodfellow2020generative} and VAE \cite{kingma2013auto}, showcasing a powerful and realistic characteristic in synthesis.
Diffusion-based motion generation also become a popular research direction, such as text2motion \cite{zhang2024motiondiffuse}. To be specific, text-guided motion can synthesize realistic and expressive results with diffusion model. In co-speech gesture synthesis, recent approaches such as GestureDiffuCLIP\cite{ao2023gesturediffuclip}, LivelySpeaker \cite{zhi2023livelyspeaker}, DiffStyleGesture \cite{ijcai2023DiffuseStyleGesture}, C2G2\cite{ji2023c2g2}, LDA \cite{alexanderson2023listen}, Remodiffuse \cite{zhang2023remodiffuse} are diffusion-based approaches.

\subsection{Semantic-aware Co-Speech Gesture Generation}
%In gesture language, semantic gesture units are crucial for conveying specific intent and improving the expressiveness of co-speech gesture.
In gesture language, semantic gesture units are crucial for conveying specific intentions, thoughts, emotions, and enhancing gestural expressiveness. 
%These units are indispensable in the field of sign language, providing a rich medium for individuals to express thoughts, emotions, and intentions clearly and profoundly.
For generating  more expressive gestures, there are also some attempts such as Gesticulator \cite{kucherenko2020gesticulator}, GestureDiffuCLIP\cite{ao2023gesturediffuclip}, LivelySpeaker{\cite{zhi2023livelyspeaker}}, Bodyformer \cite{pang2023bodyformer}.
Specifically,
%Gesticulator \cite{kucherenko2020gesticulator} and Trimodal  \cite{yoon2020speech} investigate the network structure to learn semantic gesture from textual features. 
GestureDiffuCLIP\cite{ao2023gesturediffuclip} adopts CLIP \cite{hafner2021clip} to align motion and text semantic, then use the textual feature as additional condition to control the generation process.
LivelySpeaker{\cite{zhi2023livelyspeaker}} is a two-stage framework for semantic and rhythmic gesture generation, which aim at decoupling the semantic gesture generation and rhythmic gesture generation. In semantic gesture generation, they also use a CLIP style module to align the semantic motion and semantic text. Besides, some approaches add semantic text as additional input to fuse semantic into the model, such as Rhythmic Gesticulator \cite{ao2022rhythmic}, Gesticulator \cite{kucherenko2020gesticulator}, FreeTalker \cite{yang2024freetalker} and Trimodal  \cite{yoon2020speech}. BodyFormer \cite{pang2023bodyformer} takes both low-level and high-level speech features as input and generates a sequence of realistic 3D body gestures in an auto-regressive manner. The high-level speech features represent the semantic gestures. With the development of LLMs, some approaches also use it to extract motion-related information from textual input, such as GesGPT\cite{gao2023gesgpt}, MoConVQ \cite{yao2023moconvq}. 
%However, these approaches lack generalization-ability and controllability in generating  semantic gestures. 
The proposed method uses Large Language Models (LLMs) to generate proper mapping between the text transcripts and semantic gesture units, which shows excellent performance in both generalization and controllability. 

In semantic gesture generation, Semantic Gesticulator\cite{zhang2024semantic}
is the most similar work to ours in the same period. Semantic Gesticulator\cite{zhang2024semantic} also attempts to synthesize semantically accurate gestures by introducing a LLM-based semantic gesture units retrieval from a predefined dataset. The gesture generator of Semantic Gesticulator\cite{zhang2024semantic} is based on the GPT-2 \cite{radford2019language}, which predicts the future gesture tokens in  an auto-regressive manner. Compared with their work, we use a strong diffusion-based method as fundamental gesture generator to fuse semantic gestures and rhythmic gestures.

\section{Method}
\subsection{Problem Formulation}
Given a specific audio, its corresponding text, and identity information, the proposed method can generate high-quality 3D skeletal co-speech gesture results. To be specific, considering a sample of raw audio %${A=\{a_0,a_1,...,a_N\}}$
$A$, we adopt Jukebox \cite{dhariwal2020jukebox} to extract its acoustic feature, which is pre-trained on large-scale music datasets. All of the motion sequences 
%$\mathcal{M}=\{m_0,m_1,...,m_T\}$
$\mathcal{M}$
are encoded into discrete tokens by VQVAE \cite{Neuralvqvae}. 
Details of VQVAE can be found in the appendix. 
By VQVAE encoding, a motion can be represented by a latent embedding $x_0$ 
%$\hat{e}=\{e_0,e_1,...,e_T\}$
. Then, we adopt transformer-based diffusion model as the denoising network to recover the latent embedding by removing the noise. 
The reverse denoising process of the optimized diffusion model $G$ parameterized by $\theta$ can recover the motion latent embedding $x_0$ conditioned on the speech audio sequence and identity information. With integration with LLM, it can provide accurate candidate semantic gestures for the speech transcripts.
Then, the candidate semantic gesture embedding are fused into the synthesized motion latent embedding by semantic injection module. Finally, the synthesized motion are decoded to continues 3D skeleton rotation by the pre-trained VQVAE. 

\subsection{Semantic Gesture Collection}

\begin{figure*}[tb]
  \centering
  \includegraphics[width=0.85\textwidth]{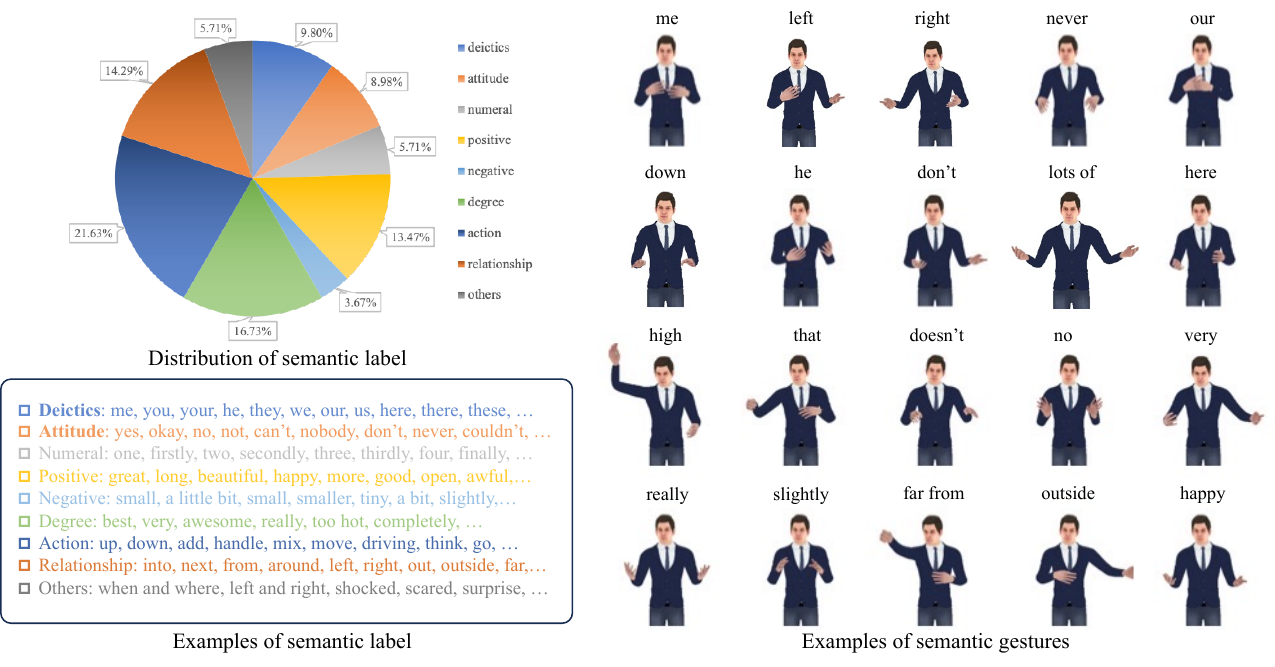}
  \caption{The statistical data of semantic gesture dataset (upper left part), some examples of semantic label (lower left part), and some semantic gesture examples (right part).}
  \label{fig:example}
\end{figure*}

Through extensive observation of numerous instances, it has been discerned that semantic gestures maintain a universality in gestural communication. Specifically, it is evident that different individuals display similar gestural expressions when articulating identical semantic concepts \cite{lascarides2009formal, zhang2024semantic}.
This observation provides a feasible methodology for the collection of an extensive array of semantic gesture actions, aimed at covering a comprehensive range of semantic contexts.
We collect a large-scale semantic gesture dataset that contains motion clips $G=\{G_0,G_1,...,G_N\}$ and corresponding text $T=\{T_0,T_1,...,T_N\}$ by manual selection and modification with the aid of technical artist. Some of these semantic gestures are borrowed from BEAT dataset. All of the semantic motion clips are shorter than 3 seconds and each motion clip is a whole independent semantic motion. After manual selection and check, we have collected 2,537 semantic motion clips. Some examples of semantic gestures and the distribution statistics are shown in Figure \ref{fig:example}. More examples can be found in the supplementary video. These motion clips are sorted into a predefined motion semantic system. 

All of semantic gesture clips are encoded into latent embedding $V_{semantic}=\{v_0,v_1,...,v_N\}$ by  a pre-trained VQVAE model. To improve the diversity of semantic gesture, we add a random noise to the embeddings during semantic injection.

\subsection{Diffusion models with semantic injection} 

SIGGesture employs a diffusion model to synthesize the latent representation of gesture, which consists of a diffusion process and a denoising process, as shown in upper part of Figure \ref{pipeline}. Given a specific real motion data distribution $p(x_0)$, our goal is learn a distribution $q_\theta(x_0)$ that is parameterized by ${\theta}$ to approximate the real distribution. 

\subsubsection{Diffusion process.}
We follow the previous DDPM \cite{ho2020denoising} definition of diffusion as a Markov noising process with latent $x_{0:T}$ that follows a forward noising process $p(x_{0:T}|x_0)$. Here, $x_0$ is the latent embedding of real motion data.  The forward diffusion process is defined as a Markov chain that gradually adds Gaussian noise to the latent representation of real motion, as follows
\begin{equation}\label{eq5}
	\begin{split}
		q(x_{0:T}|x_0) = \prod_{t=1}^{T}q(x_t|x_{t-1}),
	\end{split}
\end{equation}
where
\begin{equation}\label{eq6}
	\begin{split}
		q(x_t|x_{t-1}) = \mathcal{N}(x_t;\sqrt{1-\beta_t}x_{t-1},\beta_t{I}). 
	\end{split}
\end{equation}
$\beta_t$ are variance schedule hyperparameters that control the distribution of Gaussian noise. In the reverse denoising process, $\beta_t$ are pre-defined constant parameters. In the diffusion process, the original latent representation of motion $x_0$ progressively substituted by random noises. When $T$ approaches infinity, the distribution of $x_T$ follows a pure white noise distribution.

\subsubsection{Denoising process.}
Through diffusion process, the real data are transferred into pure white noise by adding random noise. On the contrary, the denoising process recovers the real data from pure white noise by removing the random noise. 
The unconditional reverse process of recover a gesture motion embedding can be represented by the following formula
\begin{equation}\label{eq7}
	\begin{split}
		p_\theta(x_{0:T}) = p_\theta(x_0)\prod_{t=1}^{T}p_\theta(x_{t-1}|x_{t}),
	\end{split}
\end{equation}
where
\begin{equation}\label{eq8}
	\begin{split}
		p_\theta(x_{t-1}|x_t) = \mathcal{N}(x_t;u_\theta(x_t,t),\beta_tI).
	\end{split}
\end{equation}
The intermediate state $x_t$  is sampled from the following defined distribution 
\begin{equation}\label{eq9}
	\begin{split}
		p(x_t|x_0) = \mathcal{N}(x_t;\sqrt{\Bar{\alpha_t}},(1-\Bar{\alpha_t}){I}).
	\end{split}
\end{equation}
where $\alpha_t=1-\beta_t$ and $\Bar{\alpha_t} = \prod_{k=0}^t\alpha_k$. The above explanation shows the process of unconditioned diffusion generation. For conditional content synthesis, the conditions should be injected into the diffusion generation procedure. {The details of the denoising network are shown in the upper right part of Figure \ref{pipeline}}.
In our method, the conditions contains audio $c_{audio}$ and speaker identity $c_{id}$. For simple notation,  we mark the two conditions as $c=[c_{audio},c_{id}]$. 
It should be noted that the audio feature is extracted by Jukebox \cite{dhariwal2020jukebox} instead of WavLM \cite{chen2022wavlm}. Jukebox \cite{dhariwal2020jukebox}, trained on large-scale music datasets, can focus on the rhythmic features rather than the speech semantic transcripts.
By injecting the condition $c$ into generation, the reverse process of each time step $t$ can be updated by the following formula
\begin{equation}\label{eq10}
	\begin{split}
		p_\theta(x_{t-1}|x_t,c) = \mathcal{N}(x_t;u_\theta(x_t,t,c),\beta_tI).
	\end{split}
\end{equation}
In a short summary, we firstly start denosing process by sampling $x_T$ from a pure white nose distribution $\mathcal{N}(0,I)$. Then, we iteratively denoise the latent variable $x_t$ to obtain the final results $x_0$.

\subsubsection{Training objective.}
For a speech condition $c$, we reverse the forward diffusion process by learning to estimate $p(x_t,t,c)$ approximate $x_0$ with a parameterized model for all times step $t$. We use a loss function proposed in \cite{ho2020denoising} to optimize the whole model, as following,
\begin{equation}\label{eq11}
	\begin{split}
		\mathcal{L} = {E_{x,t}[\parallel x- p(x_t,t,c) \parallel^2_2]}. 
	\end{split}
\end{equation}
During training, the model is trained under conditional and unconditional manner with a specific probability. Once the training converges, the diffusion model can predict the noise parameters by considering the unconditional setting and condition $c$.

\subsubsection{Inference with semantic injection.}

The inference process involves solely the denoising process of diffusion models. The pipeline of inference process is shown in the lower part of Figure \ref{pipeline}.
%The semantic gesture is injected into the latent variable $x_t$ during the generation. 
To be specific, a speech with its corresponding text will be fed into the generation pipeline. It should be noted that the words in the text are aligned with speech in the timeline. Each word has a duration time. Then, the texts are fed into the LLMs to produce candidate semantic gestures $G_{candicate}=(g_0,...,g_N)$ with a pre-defined prompt. $N$ is the number of semantic gestures in this text. The interaction with LLMs process is shown in the left-top corner of inference pipeline in Figure \ref{pipeline}. With the ability of LLMs, the proposed method can process various languages, such as Chinese and Japanese.  

The candidate semantic gestures have been encoded into latent embedding. We set a control $K$ to determine the degree of semantic injection. From time step $T$ to $K$, the denoising latent variable is marked as $\hat{x_t}$ that contains the semantic information. 
\begin{equation}\label{eq12}
	\begin{split}
		p_\theta(x_{K:T}) = p_\theta(x_T)\prod_{t=K}^{T}p_\theta(x_{t-1}|\hat{x}_{t}).
	\end{split}
\end{equation}
The semantic gesture $G_{candidate}$ can be injected into the latent variable $x_t$ by the follow formula
\begin{equation}\label{eq13}
	\begin{split}
		\hat{x}_t = x_t * m + G_{candidate}*(1-m),
	\end{split}
\end{equation}
where $m$ denotes the timeline mask of semantic words. From time step $K$ to 1, the denoising latent variable $x_t$ does not contain semantic injection step,  which can preserve the structure and smoothness of the generated gesture and improve the diversity of the results. 

\subsubsection{Generating long sequence.}
For synthesizing long motion sequence, the common practice uses auto-regressive manner, which generates content sequentially. However, generative models trained on small pieces may cause unnaturalness and defectiveness. Instead of auto-regressive manner, we adopt DiffCollage \cite{zhang2023diffcollage} to generate long motion sequence. DiffCollage is a scalable probabilistic model that can synthesize large content in parallel. Specifically, we denote a long sequence as ${\boldmath{m} = [m^0, m^1,m^2]}$, where $m^2$  is the out-painted sequence generated by the conditional model $m^2 | m^1$. Notably, this procedure makes a conditional independence assumption that $q(m^2|m^0,m^1)=q(m^2|m^1)$.
\begin{equation}\label{eq111}
	\begin{split}
		q(m) = q(m^0, m^1, m^2) = q(m^0, m^1)q(m^2|m^1).
	\end{split}
\end{equation}
Further, we can obtain the following formula,
\begin{equation}\label{eq112}
	\begin{split}
		q(m) = \frac{q(m^2, m^1)q(m^1, m^0)}{q(m^1)}. 
	\end{split}
\end{equation}
The score function of $q(m)$ in diffusion model can be represented as a sum over the scores of shorter gesture motions.
\begin{equation}\label{eq113}
	\begin{split}
		\nabla q(m) = \nabla q(m^2, m^1) + \nabla q(m^1, m^0) - \nabla q(m^1). 
	\end{split}
\end{equation}
From formula \ref{eq113}, it can be found that we can  synthesize different gesture clips in parallel since all individual scores can be calculated independently.
Then, then we can merge the scores to compute
the score of the target long audio condition. 
For more technical details, please refer to DiffCollage \cite{zhang2023diffcollage}.

\subsection{Large-Scale Motion Corpus Collection}
In 3D animation, high-quality data collection is remarkably expensive, professional, and time-consuming. High-quality motion capture requires specialized cameras, suits, and markers. Besides, it often requires experienced technicians, actors, and animators to set up, perform, and process the captured data. The process of capturing, cleaning, and integrating motion data is labor-intensive and time-consuming, contributing to the overall expense. In recent years, several high-quality datasets are released to the research community, as shown in Table \ref{datasets}. BEAT \cite{liu2022beat} is the largest dataset for co-speech gesture synthesis, which contains 76 hours data. However, data-driven approaches with larger models are also short for data. In the inspiration of large-language training paradigm, noise data for pre-training and high-quality data for fine-tuning, we aim at collecting a large-scale gesture corpus for pre-training by video motion capture technique such as HybridIK \cite{li2021hybrik}, SMPLify \cite{bogo2016smplify}. To this end, we collect a large mount of videos about speech or talk from TED \footnote[1]{\url{https://www.ted.com}} and YiXi \footnote[2]{\url{https://yixi.tv}}. Then, we conduct the data post-processing pipeline to obtain the final shots of interests. Table \ref{tab:data} shows the statistical data of the motion corpus collection. Finally, we obtain about 400 hours 3D gesture assets of speech. As shown in Table \ref{datasets}, the proposed dataset is significantly larger than previous datasets by at least 6 times.
%More details can be found in supplementary material. 
\begin{table}[tb]
  \caption{Statistical data of the collected motion corpus collection.
  }
  \label{tab:data}
  \centering
  \setlength{\tabcolsep}{1.4mm}{
  \begin{tabular*}{1.0\linewidth}{llll}
\toprule
Items & Number & Average length & Total length\\
\midrule
original videos & 6,032 & $\sim$23min & $\sim$2,312h\\
shots-of-interests & 138,209&$\sim$20s& $\sim$746h\\
final results &70,784 &$\sim$20s &$\sim$393h\\
  \bottomrule
  \end{tabular*}}
\end{table}

\begin{figure*}[tb]
  \centering
  \includegraphics[width=1.0\textwidth]{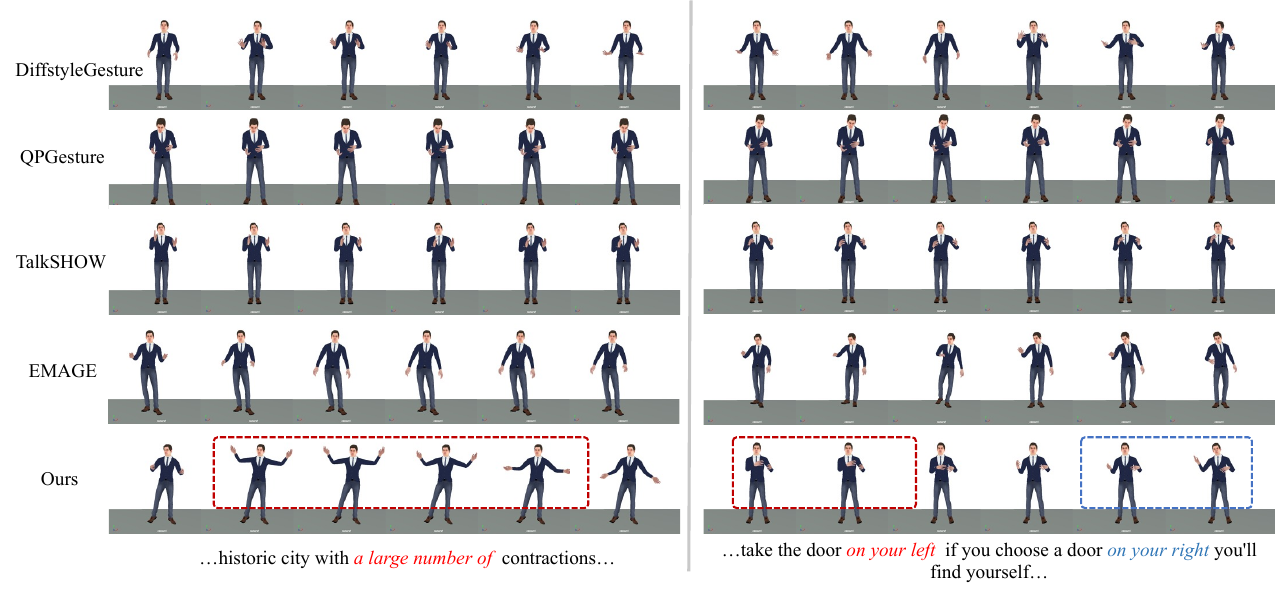}
  \caption{Visualization results between the proposed method and other state-of-the-art methods.}
  \label{fig:compare}
\end{figure*}

\section{Experiments}
\subsection{Datasets}
In this paper, we use the largest high-quality speech-gesture dataset BEAT to evaluate the proposed method. BEAT is constructed by using commercial motion capture system, including face emotion and body motion.  It contains about 76 hours motion-audio paired data of 30 speakers talking about different topics. 
We thoroughly post-process the dataset by removing some defective motion sequences such as speaker drinking, wandering, and incorrect motion.
Then, the processed BEAT dataset contains 46 hours motion data and its corresponding text and audio.
%In our experiments, we only use the body motion data (rotation angle of a joints). 
During training the VQVAE model, the original Euler angles are converted to rotation matrix for better convergence. During training the diffusion-based generation model, all motions are represented by latent embeddings in VQVAE latent space. The datasets collected from Internet by video motion capture are only used for pre-training the diffusion model because of its relatively low quality.

\subsection{Evaluation Metrics}
For content generation tasks, human objective evaluation is the most important evaluation method because of lacking clear ground and explicit criteria. Therefore, the human objective evaluation is the main evaluation method in our experiment.
Each motion slice is re-targeted to Mixamo\footnote[3]{\url{https://www.mixamo.com}} model and rendered as a video. 
The evaluators are asked to rate the slices from the following four aspects respectively: (1) Naturalness (Nat.), (2) Rhythm (Rhy.), (3) Diversity (Div.), (4) Semantic (Sem.). The scores assigned to each rating are in the range of 1-10, corresponding from worst to best. All scores are normalized to 0-1. Besides, we also calculate some quantitative metrics adopted by previous works.
FGD \cite{yoon2020speech} measures the distribution difference between generated data and ground truth. Beat Consistency (BC) \cite{li2021ai} calculates the average distance between every audio beat and its nearest motion beat.
Diversity is a metric of measuring the variations of the synthesized gesture.
Semantic-Relevant Gesture Recall (SRGR) \cite{liu2022beat} measures the semantic correct key point of synthesized data by comparing it with the ground truth.

\subsection{Implementation Details}
The whole training process consists of two steps. Firstly, we train the VQVAE model by using the Adam optimizer with learning rate 0.001 for 400 epochs. Then, we train the diffusion-based gesture generation model using the Adam optimizer with learning rate 0.0002 for 3000 epochs with batch-size 256. The number of diffusion steps for training and inference is set as 1000. During the training of diffusion model, we firstly use the collected data for pre-training (2000 epochs with batch-size 256), and then use the high-quality dataset for fine-tuning (1000 epochs). In the diffusion model, the variances $\beta_t$ are predefined to increase linearly from 0.0001 to 0.02, with a total of 1000 noising steps set as $T = 1000$.
The length of the input audio feature is 12 seconds. We train the diffusion model on 32 V100 GPUs for about one week.

\subsection{Comparing Baselines}
We compare our method on BEAT dataset with several representative state-of-the-art methods in the last year, including   
DiffuseStyleGesture \cite{ijcai2023DiffuseStyleGesture}, 
%LDA\cite{alexanderson2023listen}, 
TalkSHOW \cite{yi2023talkshow}, 
QPGesture \cite{yang2023qpgesture},  
EMAGE \cite{liu2023emage}. 
DiffuseStyleGesture \cite{ijcai2023DiffuseStyleGesture} is a diffusion model-based co-speech gesture synthesis approach, which takes the motion style into generation process. 
TalkSHOW \cite{yi2023talkshow} uses a simple full-body speech-to-motion generation framework with auto-regressive manner, in which the face, body, and hands are modeled separately.
QPGesture \cite{yang2023qpgesture} is a quantization-based and phase-guided motion-matching approach for co-speech gesture synthesis.
EMAGE \cite{liu2023emage} leverages a masked gesture reconstruction to enhance audio-conditioned gesture generation.

\begin{figure*}[tb]
  \centering
  \includegraphics[width=0.95\textwidth]{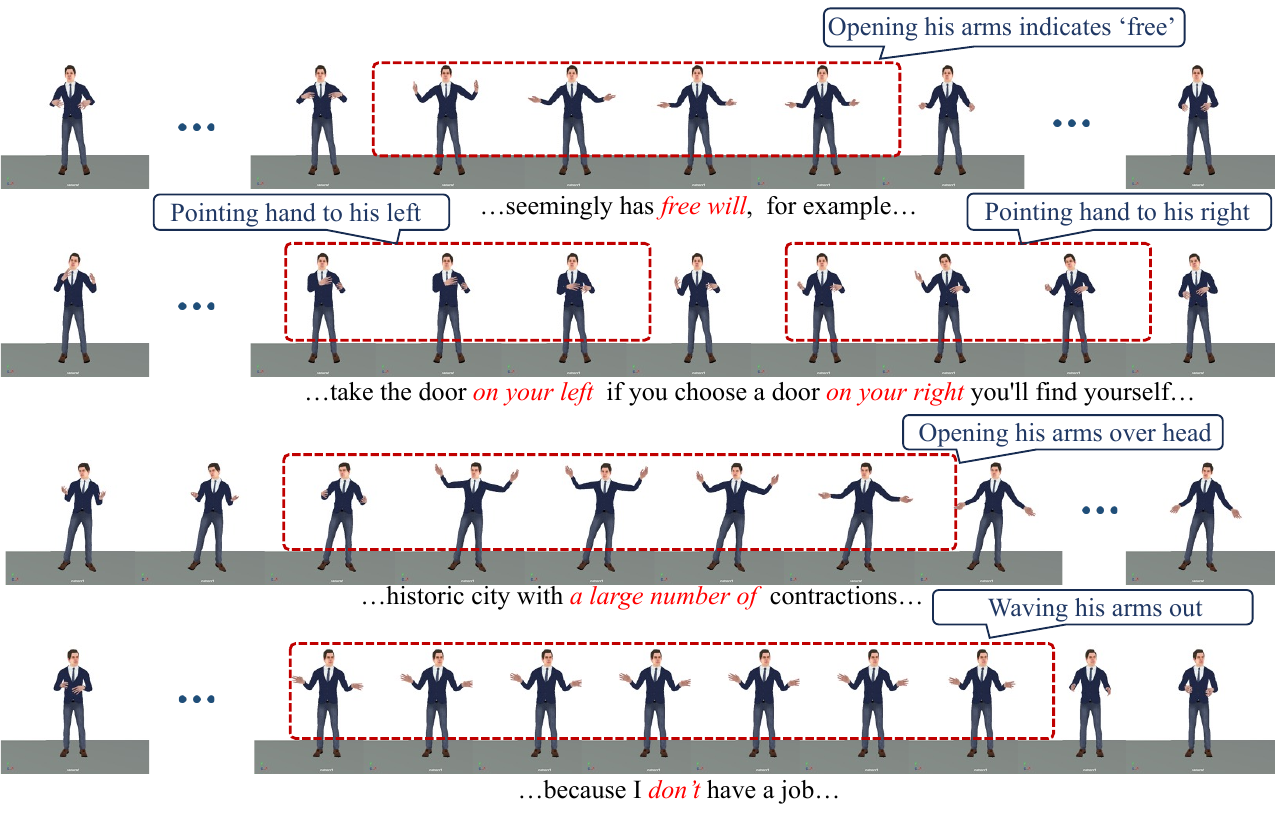}
  \caption{The visualization results of the proposed method. The gestures in the red box are semantic gesture, which is labeled by the content in green box.}
  \label{fig:visual}
\end{figure*}

\subsection{Evaluation Results}

\subsubsection{ Qualitative Results}
For this task, there is lack of absolutely objective metric to evaluate the performance of various methods. Therefore, we strongly recommend the readers to view our demonstration video to gain an intuitive understanding of the qualitative outcomes. Figure \ref{fig:compare} shows some synthesized results of these baselines conditioned on the same speech. From these results,  we can observe that the synthesized co-speech gestures of the proposed method are more realistic, agile, and diverse than those of the baselines on BEAT datasets. The baselines suffer different extents of jittering in the results, especially lacking of semantic gesture and naturalness. DiffuseStyleGesture \cite{ijcai2023DiffuseStyleGesture} is also a diffusion-based method, which directly learns the origin motion representations. However,its results lack  smoothness and naturalness (e.g. foot sliding). QPGesture \cite{yang2023qpgesture} exhibits noticeably slower and less varied motion, while ours demonstrates agility comparable to actual movement. For the specific semantic expressions, apart from our approach, other methods cannot effectively synthesize proper semantic gestures.
More results are shown in Figure \ref{fig:visual}. The results indicate that our method can generate proper semantic gestures for the speech semantics. For example, "free will" refers to the first row with open arms; "left" and "right" indicate pointing to the corresponding direction. These results show the surpassing capability of SIGGesture in synthesizing semantic accurate and natural gestures.

\subsubsection{User study.} 
As we all known, the evaluation of generative task is very subjective. Although some metrics are introduced to evaluate the performance, there is a huge gap between metric evaluation and human visual perception. 
We conduct a user study to evaluate the performance of different baselines. Specifically, we randomly select 50 synthesized samples and shuffle results of different methods. 
Following previous works\cite{zhi2023livelyspeaker,ijcai2023DiffuseStyleGesture}, the participants are asked to scoring the gesture based on naturalness, rhythm appropriateness, diversity, and semantic consistence for the shuffled visual results. 
%For each aspect, the participants should give a point (from 0 to 10) for the shuffled visual results. 
%The details of user study can be found in supplementary materials.
It's important to note that, for diversity, we enable users to score motion diversity under the condition of smoothness and naturalness. 
The results of user study are shown in Table \ref{tab:user}. The results show that the generated gestures of SIGGesture are dominantly preferred on four metrics over the comparing baselines. Especially, SIGGesture exceeds these baselines by a large margin in semantic relevancy.
%The SIGGesture can achieve competitive result with EMAGE in term of diversity. 
It can be concluded that large-scale pre-training benefits rhythm and diversity, while semantic injection is advantageous for semantic consistency.
Overall, SIGGesture is capable of generating more realistic, synchronized, diverse, and understandable gestures that humans prefer.

\begin{table}[tb]
  \caption{The statistical data of user study by comparing the proposed method to baselines. The "$\downarrow$" means the lower, the better.  The "$\uparrow$" means the higher, the better.
  }
  \label{tab:user}
  \centering
  \setlength{\tabcolsep}{2.2mm}{
  \begin{tabular*}{1.0\linewidth}{lllll}
\toprule
Methods & Nat.$\uparrow$ & Rhy.$\uparrow$ & Div.$\uparrow$&Sem.$\uparrow$\\
\midrule
%CaMN &0.23 & 0.30 & 0.53&0.29\\
%DiffGesture & 2 &&&\\
DiffuseStyleGesture & 0.78&0.82&0.65&0.57\\
TalkSHOW &0.56&0.63&0.57&0.37\\
QPGesture &0.62 &0.72&0.53&0.56\\
EMAGE & 0.80 &0.81&{0.79}&0.60\\
\textbf{SIGGesture}&\textbf{0.82}&\textbf{0.87}&\textbf{0.83}&\textbf{0.92}\\
  \bottomrule
  \end{tabular*}}
\end{table}

\begin{figure*}[tb]
  \centering
  \includegraphics[width=1.0\textwidth]{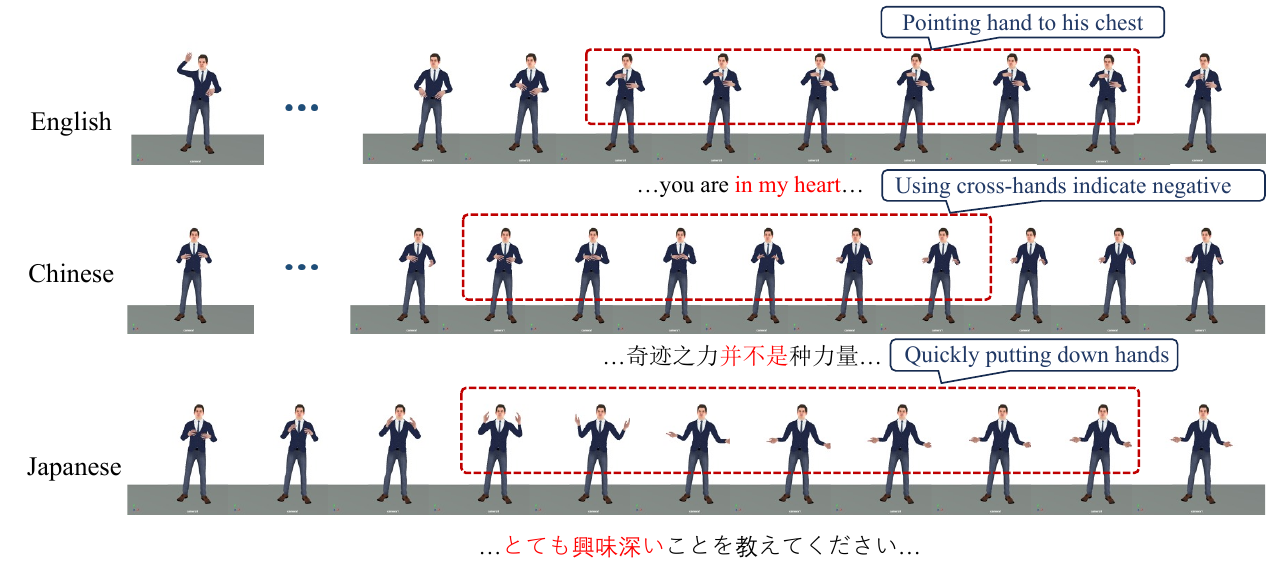}
  \caption{The visual results of the proposed method on in-the-wild speeches (English, Chinese, and Japanese). The gestures in the red box are semantic gesture, which is labeled by the content in green box. }
  \label{fig:generalization}
\end{figure*}

\subsubsection{Quantitative Results.} 
We compare our method with the baselines using four metrics on BEAT dataset. The comparison results are shown in Table \ref{tab:quantitative}. As discussed in recent works \cite{tseng2023edge, ao2023gesturediffuclip}, these metrics have obvious weakness because these valued can not keep consistent with the visual results. For example, TalkSHOW \cite{yi2023talkshow} generates unnatural body movements based on the given audio, as shown in our supplementary video, but the results of these metrics can not show this gap. Besides, the SRGR can not show the difference of various methods, because SRGR only considers the ground truth semantic motion for a specific speech transcripts. 
Therefore, the SRGR of our method is lower than TalkSHOW. The reason is that this metric cannot accurately represent semantic performance. Because SRGR calculates the similarity (distance of skeleton) between the generated motion and the GT motion (with semantic labels and duration) in 3D space. For the same sentence, different semantic words can be expressed.
All the same, the proposed method can outperform previous works across most metrics. Although we only consider two modalities (audio and speaker identity), the proposed method can excels state-of-the-art model that employs all five modalities, such as DiffStyleGesture.

\begin{table}[tb]
  \caption{The statistical data of quantitative results by comparing SIGGesture to the baselines. The "$\downarrow$" means the lower, the better.  The "$\uparrow$" means the higher, the better.
  }
  \label{tab:quantitative}
  \centering
  \begin{tabular*}{1.0\linewidth}{lllll}
\toprule
Methods & FGD$\downarrow$ & SRGR$\uparrow$ & BC$\uparrow$&Diversity$\uparrow$\\
\midrule
DiffuseStyleGesture  & 10.14&0.233&0.504&11.975\\
TalkSHOW &7.313&\textbf{0.279}&0.463&12.859\\
QPGesture&19.921 & 0.209& 0.453&9.438 \\
EMAGE &5.430&0.272&0.679&13.075\\
\textbf{SIGGesture}&\textbf{2.021}&0.263&\textbf{0.707}&\textbf{14.020}\\
  \bottomrule
  \end{tabular*}
\end{table}

\subsection{Ablation study}
In this subsection, we evaluate the contribution of two main parts, large-scale pre-training and semantic injection, by user study. The comparison results are shown in Table \ref{tab:allation}. The findings reveal that the pre-training with noise data can improve the quality of basic rhythmic gesture. In addition to pre-training, the proposed semantic injection can significantly improve the quality of semantic gesture. The experimental results align with our hypothesis that pre-training with large-scale data can enhance the fundamental  capabilities of synthesizing rhythmic gestures, and incorporating semantic injection with LLMs can accurately generate semantic gestures.
\begin{table}[tb]
  \caption{The statistical results of ablation study. "w/o" means "without".
  }
  \label{tab:allation}
  \centering
  \setlength{\tabcolsep}{2.2mm}{
  \begin{tabular*}{1.0\linewidth}{lllll}
\toprule
Methods & Nat.$\uparrow{}$ & Rhy.$\uparrow$ & Div.$\uparrow$&Sem.$\uparrow$\\
\midrule
baseline &0.56 &0.63&0.50&0.38\\
w/o pre-training & 0.64&0.72&0.72&0.87\\
w/o semantic injection&0.81&0.85&0.76&0.53\\
\textbf{SIGGesture (full)} &\textbf{0.82}&\textbf{0.87}&\textbf{0.83}&\textbf{0.92}\\
  \bottomrule
  \end{tabular*}}
\end{table}

\subsection{Generalization}
To evaluate the generalization ability, we test the proposed method on a broader set of audio samples, including out-of-domain English, Chinese and Japanese.
The results are shown in Figure \ref{fig:generalization}.
It is also recommended to view the supplementary video for more intuitive understanding.  
As shown in the visualization, the proposed method can accurately capture the key semantics and present precise gestures cross different language. The enhanced generalization capacity of our methodology is attributable to two principal components. Firstly, the employment of pseudo-labeled data can significantly improve the generalization ability of rhythmic gestures. Secondly, disentangling the generation of rhythmic and semantic gestures by semantic injection can strengthen the controllability of semantic gesture synthesis. 
We have established an extensive collection of semantic gestures, encompassing a comprehensive range of gestural expressions.
We leverage LLMs as a bridge to link the semantic gesture database with linguistic contexts, which enables robust generalization across diverse languages. 
This method effectively addresses the challenges posed by the sparsity and long-tail distribution of semantic gestures. As a result, the proposed method has distinguished generalization over different speech inputs.

\section{Conclusion}
In this paper, we introduce a novel method called SIGGesture for high-quality co-speech gesture synthesis. Our approach leverages a semantic injection technique using LLMs to enhance controllability, interpretability, and generalization. Thanks to the capabilities of LLMs, our method can be seamlessly applied to other languages without requiring any complex adjustments. Additionally, we have developed a large-scale 3D gesture dataset sourced from Internet videos to pre-train our diffusion-based synthesis model. By employing a pre-training and fine-tuning paradigm, our generated gestures exhibit greater robustness to variations in audio input. The proposed datasets and accompanying statistical experiments aim to advance the field of co-speech gesture synthesis, including areas such as controllable gesture synthesis and gesture foundation models. Looking ahead, our research will focus on generating full-body animations with enhanced and detailed expressiveness.

\section{Acknowledgements}
We thank Wenhao Ge and Baocheng Zhang for their kindly support.

\bibliographystyle{ACM-Reference-Format}
\bibliography{final_arxiv.bib}

\clearpage
\appendix
\section{Motion Representation}
\begin{figure}[tb]
  \centering
  \includegraphics[width=0.45\textwidth]{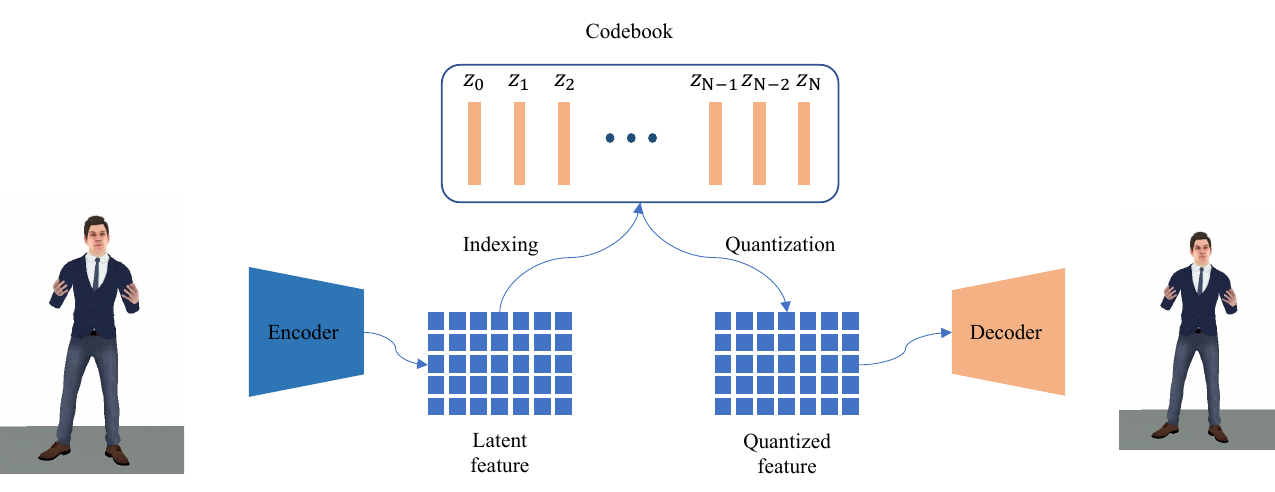}
  \caption{The architecture of VQVAE. It contains an encoder, a decoder and a codebook.}
  \label{fig:vqvae}
\end{figure}

The diffusion model synthesizes the latent feature, and then the decoder of the Vector Quantized Variational Autoencoder (VQVAE) recovers the skeletal rotation from the latent space. VQVAE \cite{Neuralvqvae} has shown excellent performance in reconstructing complex content, including image, animation, video, etc. This section will present the details of the VQVAE model. We use high-quality skeletal rotations as motion representations and parameterize these rotations using rotation matrices relative to a reference T-pose. The original motion is denoted as $\mathcal{M}=\{m_0,m_1,...,m_L\}$, ${m_i \in R^{J\times 9}}$, where ${J}$ is the number of skeletal bones and 9 is the dimension of rotation matrix.
The initial FPS of different datasets is converted to 30 FPS for a unified standard. 
For a specific motion sequence $\mathcal{M}$, we use VQVAE to encode and quantize it into a finite codebook $\mathcal{Z}=\{z_i\}_0^N$, where ${N}$ is the size of codebook and ${z_i}$ is a specific lexeme. 
By using VQVAE, we can quantize the continuous motion into discrete tokens marked as $\mathcal{M}_{token}=\{\mathcal{T}_{0},...,\mathcal{T}_{L/4}\}$ with 4$\times$ down-sampling. 
The framework of VQVAE is shown in Figure \ref{fig:vqvae}.
A piece of original motion $\mathcal{M}$ firstly is converted to a latent embedding $\hat{e}=\{e_0,e_1,...,e_{L/4}\}$. Then, ${\hat{e}}=\{e_0,e_1,...,e_{L/4}\}$ is substituted by its nearest vector ${z_j}$ in the codebook, which can be denoted as $\arg \min \limits_j \parallel e_i - z_j\parallel$.

The encoding process can be represented by the following formula
\begin{equation}\label{eq1}
	\begin{split}
		\hat{e}_i = VQ_{encoder}(\mathcal{M}).
	\end{split}
\end{equation}
The decoder will reconstruct the motion $\mathcal{\hat{M}}$ from the quantized latent feature as the following formula
\begin{equation}\label{eq2}
	\begin{split}
		\hat{\mathcal{M}} = VQ_{decoder}(\hat{e}_i).
	\end{split}
\end{equation}
$VQ_{decoder}$ and $VQ_{decoder}$ are modelled by Convolutional Neural Networks.
In the training process, the encoder, the decoder, and the codebook are optimized by the following loss function
\begin{equation}\label{eq3}
	\begin{split}
		\mathcal{L}_{VQ} = \mathcal{L}_{recon} +  \parallel \hat{e}-sg(e)\parallel+\parallel sg(\hat{e})-e\parallel,
	\end{split}
\end{equation}
where $\mathcal{L}_{recon}$ is the reconstruction error and $sg[*]$ denotes the stop gradient operator. The reconstruction error is calculated by the following formula
\begin{equation}\label{eq4}
	\begin{split}
		\mathcal{L}_{recon} = \parallel \mathcal{\hat{M}} - \mathcal{M}\parallel + \alpha_0\parallel \mathcal{\hat{M}} - \mathcal{M}\parallel + \alpha_1\parallel \mathcal{\hat{M}}^{\prime\prime} - \mathcal{M}^{\prime\prime}\parallel,
	\end{split}
\end{equation}
where ${\mathcal{M}^{\prime}}$ and $\mathcal{M}^{\prime\prime}$ are the velocity and acceleration of the original motion sequence $\mathcal{M}$ respectively.
All motions are encoded into latent space for training the diffusion model. In our pipeline, VQVAE serves as a fundamental model used to encode motions into latent space and to recover motions from latent space. 

\section{User Study}
We place significant importance on the effectiveness and fairness of our user study evaluations. To this end, we have recruited 30 volunteers (15 men and 15 women), including 5 researchers with expertise in gesture or animation synthesis.
Each participant evaluates 250 co-speech gesture videos, which are generated from 50 audio clips using 5 different methods. To ensure that participants understand each metric and can accurately differentiate between them, we provide detailed instructions and training. This training covers the standard evaluation process and explains the differences between the various metrics. The detailed difference of these metric are as following.

The user study focus on the following aspects, Naturalness, Rhythm, Diversity, Semantic. Naturalness refers to how realistically and convincingly the generated motion movements mimic real-world behavior and physics. This includes evaluating whether the movements are fluid, smooth, and consistent with real movements and real-world physical laws. 
Rhythm metric typically evaluates how well the visual animation aligns with the rhythm or timing of the accompanying audio. This involves examining whether the gesture movements (mainly are hand movements) synchronize effectively with the beats or tempo of the audio.
Diversity metric measures how much variety is present in the generated gesture animations produced in response to a specific audio. A higher score of diversity ensures that the synthesized co-speech gestures are not only synchronized with the audio but also rich and varied in a range of different movements and styles rather than repeating similar patterns or showing minimal variation.
Semantic metric evaluates how well the animations convey the intended meaning, specific semantic or emotional content of the audio. A high Semantic score indicates that the generated gesture accurately interprets and represents the content of the audio. 

Participants first browse the videos to familiarize themselves with the content. They then formally score the videos, with 5 animations from different methods displayed simultaneously on the screen. The differences in scores will reflect the performance variations among the different algorithms.

\section{Video-based Collected dataset}
\begin{figure}[tb]
  \centering
  \includegraphics[width=0.45\textwidth]{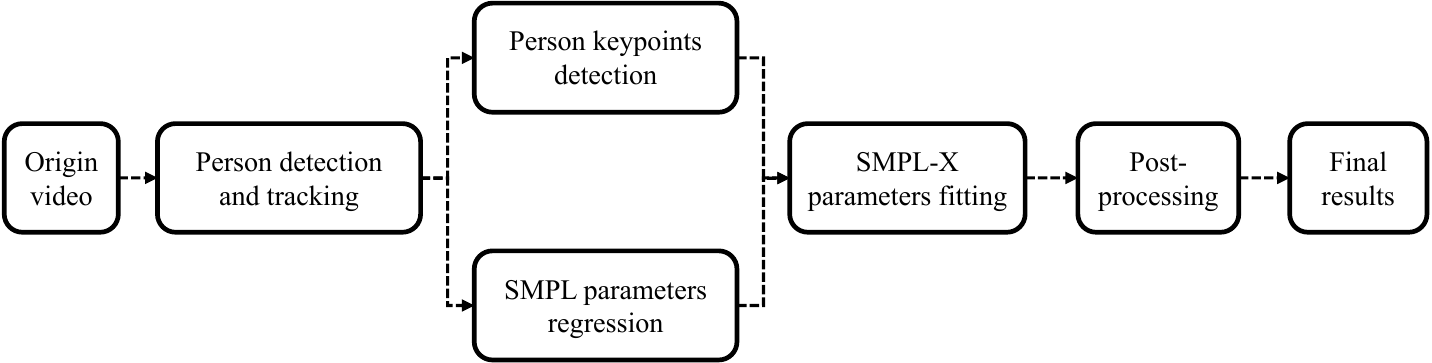}
  \caption{The process pipeline of video-based gesture motion capture. }
  \label{fig:process}
\end{figure}

To ensure safety, raw data is collected from official TED and YiXi sources. Each video on TED and YiXi varies in terms of speaker’s gender, age, speech topics, and talking styles. This diversity enriches the dataset with a variety of gesture types. Finally, the dataset contains about 50\% English and 50\% Chinese. Following previous work \cite{yoonICRA19},
the selection criteria for shots of interest include: containing speech content, having a stable background, displaying the entire upper body, facing the camera, and ensuring the body occupies at least 50\% of the screen. These video clips are then processed through a gesture motion capture pipeline, depicted in Figure \ref{fig:process}, which comprises five key components. Person detection utilizes Yolo-v8 \cite{yolov8_ultralytics}, while tracking employs the Sort algorithm \cite{Bewley2016_sort}. \cite{Bewley2016_sort}. Person key points are detected by using ViTPose \cite{xu2022vitpose}, which are trained on a large-scale dataset. SMPL parameters fitting is like SMPLify \cite{bogo2016smplify}, which can boost the pose accuracy of SMPLX \cite{SMPL-X:2019}. Post-processing part mainly contains motion smoothing and quality checking. 

Finally, the data is represented using the NEUTRAL SMPL-X skeleton. It took about 2 months and utilized 64 V100 GPUs for collecting the video-based gesture data.

\section{LLM prompt}
The prompt is divided into three components: the task definition, a series of examples for few-shot learning, and the final task.  
During inference, the LLM is employed solely to generate accurate semantic gesture mappings from a predefined set. Our experiments utilize GPT-4 \cite{gpt4} to perform this task and also involve a comparison of various prompting strategies, such as zero-shot and few-shot learning. In our findings, zero-shot learning prompts generally fail to produce appropriate semantic gestures. Conversely, few-shot learning prompts demonstrate effectiveness in performing the annotation task and yield meaningful results. It should be noted that these examples for few-shot learning contain both positive and negative examples. Below, we provide an example of the prompt, noting that some terms have been simplified due to page constraints.

\begin{figure}[tb]
  \centering
  \includegraphics[width=0.45\textwidth]{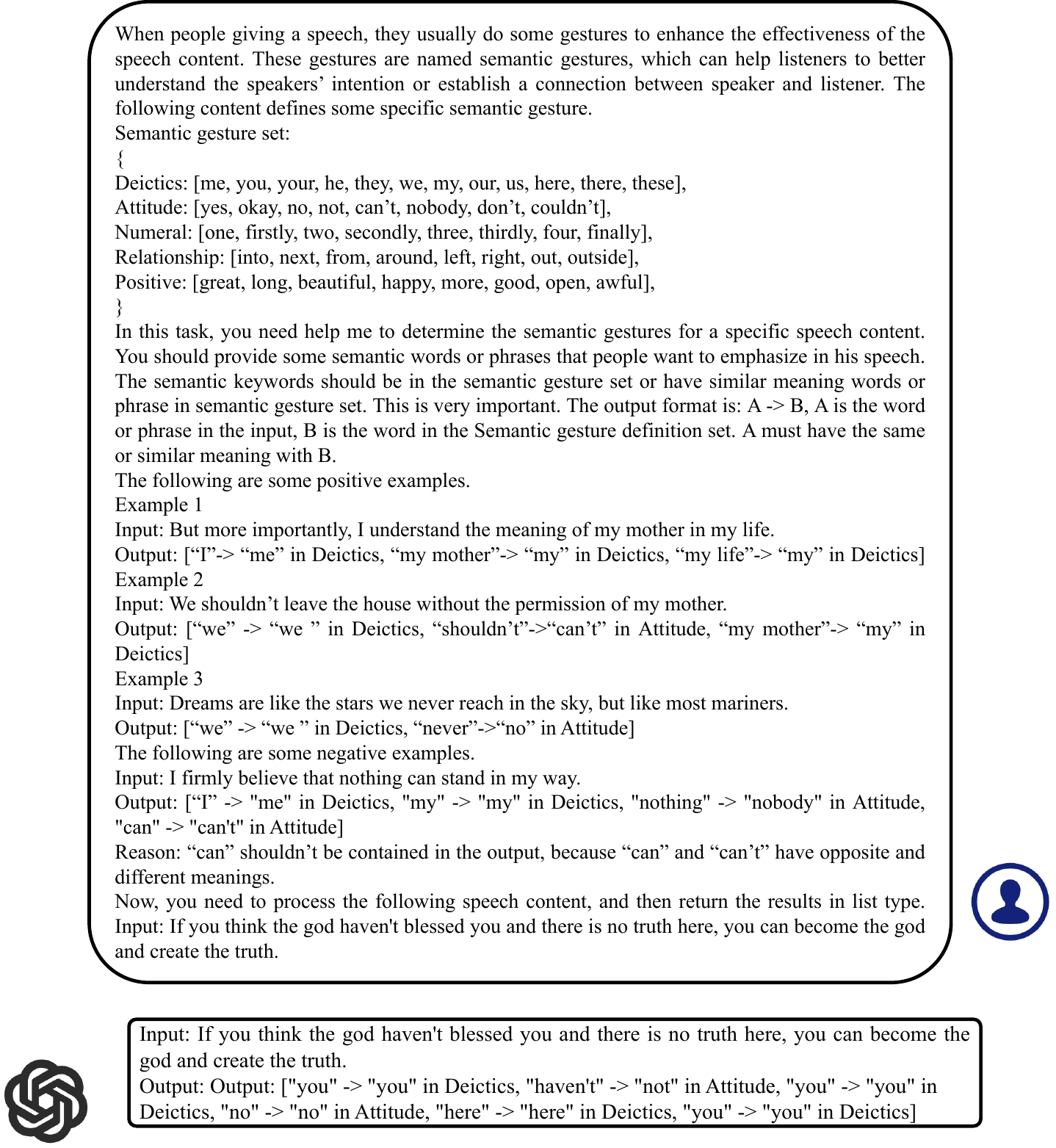}
  \caption{An example of the LLM prompt for generating proper candidate semantic gesture. }
  \label{fig:prompt}
\end{figure}

\end{document}